\documentclass[aps,prd,twocolumn,groupedaddress,nofootinbib,showkeys,showpacs,numbers,sort&compress]{revtex4}

\usepackage{graphicx}
\usepackage{dcolumn}
\usepackage{bm}

\newcommand*{\no}{\noindent}
\newcommand*{\bea}{\begin{eqnarray}}
\newcommand*{\eea}{\end{eqnarray}}
\newcommand*{\be}{\begin{equation}}
\newcommand*{\ee}{\end{equation}}

\newcommand*{\pref}[1]{(\ref{#1})}

\newcommand*{\nn}{\nonumber}

\begin{document}


\title{$\rho$-meson, Bethe-Salpeter equation, and the far infrared}

\author{Martina Blank}
\author{A. Krassnigg}
\author{Axel Maas\footnote{Present address: Institute for Theoretical Physics, Friedrich-Schiller-University Jena, Max-Wien-Platz 1, D-07743 Jena, Germany}}

\email[]{axelmaas@web.de}
\affiliation{Institute of Physics, Karl-Franzens-University Graz,
	    Universit\"atsplatz 5, A-8010 Graz, Austria}

\date{\today}

\begin{abstract}
The Bethe-Salpeter equation in QCD connects the gauge-dependent gluon and quark degrees of freedom
with the gauge-invariant properties of mesons. We study the $\rho$ meson mass and decay constant for various
versions of the gauge-dependent input functions discussed in the literature, which start to differ generically
below the hadronic scale, and show qualitative different infrared behavior. We find that, once the gauge-dependent
quark-gluon vertex is permitted to vary as well, the $\rho$ mass and decay constant is reproduced equally well
for all forms investigated. A possible conclusion from this is that these $\rho$-meson properties are only
sensitive to changes in the input at scales above a few hundred MeV.
\end{abstract}


\maketitle

\section{Introduction}

In the context of the Dyson-Sch\-wing\-er equations (DSEs) of QCD, the homogeneous Bethe-Salpeter equation (BSE) has been quite
successfully used to study mesonic properties \cite{Alkofer:2000wg,Fischer:2006ub,Roberts:2007jh,Krassnigg:2009zh},
with a natural extension to baryons \cite{Eichmann:2007nn,Nicmorus:2008vb,Eichmann:2009qa}. Such model calculations require both a numerical treatment as well as a truncation of the infinite coupled tower of DSEs, the latter of which may introduce artifacts in the results. Since the high-momentum behavior of correlation functions is essentially under control by perturbation theory, such artifacts will be most problematic at low momenta, i.\ e., at the order of the hadronic energy scale, or below. Unfortunately, this is also the energy domain which is most important for hadronic properties. Here, we investigate whether the whole low-energy domain is relevant, or only a part of it. The aim of this study is thus to investigate in particular to which extent the far infrared properties are of relevance. This investigation is performed for the combined properties of the $\pi$ and $\rho$ mesons, in particular, their masses and decay constants as a straight-forward and instructive indicator for the effects under consideration.

The reason to suspect that only a certain energy window, between a few hundred MeV and a few GeV, is relevant stems from two sources. On the one hand, indications have been found from various sources \cite{Glozman:2009cp,Glozman:2010zn,Bhagwat:2006pu} that at least for some mesons only the energy domain around a half to one fermi is significantly contributing to their properties. On the other hand, recent results on the gauge-dependent correlation functions of the elementary constituents of mesons, the quarks and gluons, has given rise to the hypothesis that the far infrared behavior may be qualitatively dominated by a gauge choice \cite{Fischer:2008uz,Maas:2008ri,Maas:2009se}, as a consequence of the Gribov-Singer ambiguity. This would again imply that the properties of mesons should be rather insensitive to the far infrared behavior, in particular of the elementary propagators and vertices.

In fact, even the precise determination of such a simple correlation function as the gluon propagator for a single fixed gauge turned out to be a rather complicated problem. Therefore, this whole problem set obtains also a different, purely technical, perspective: If the mesons are indeed rather insensitive to the far infrared, can then an incorrect input still yield the correct meson properties, and thus fool us into believing that we have obtained the correct input?

For the present purpose two qualitatively different inputs are chosen, the so-called scaling-type \cite{Fischer:2008uz} and
decoupling-type \cite{Fischer:2008uz,Binosi:2009qm,Dudal:2009xh,Boucaud:2010gr} solutions for the Yang-Mills sector
of Landau-gauge QCD. While the former shows an infrared power-law behavior, characterized by scaling relations between the different
correlation functions \cite{Zwanziger:2003cf,Lerche:2002ep,Alkofer:2004it,Fischer:2009tn}, the latter exhibit
screening \cite{Fischer:2008uz,Binosi:2009qm,Dudal:2009xh,Boucaud:2010gr}. At small volumes on a lattice, it is found that scaling-type and decoupling-type behavior can be associated with how the non-perturbative Gribov-Singer ambiguity is treated \cite{Maas:2008ri,Maas:2009se,Fischer:2007pf,Fischer:2005ui}. However, at the present time it is completely unclear whether this persists in a continuum and infinite-volume formulation.

The differences between both solution starts to manifest itself only at energies at or below a few hundred MeV, and becomes substantial only at even lower scales. Thus, they provide an ideal laboratory to study the questions raised above. It is possible to take various different points of view with respect to these questions:
\begin{enumerate}
 \item If the lattice results \cite{Glozman:2009cp,Glozman:2010zn} are correct, it should not matter which of the solution is taken, since the
 $\rho$-meson's largest gauge-invariant contribution appears to be only of the order of half a fermi large.
 \item Since it has been argued that only one of the solutions remains in the infinite-volume limit \cite{Binosi:2009qm,Dudal:2009xh,Boucaud:2010gr} (which would then most likely be of decoupling-type \cite{Cucchieri:2008fc,Bogolubsky:2009dc}), it can be tested whether meson properties would be at all able to distinguish between the solutions. In particular, since this implies at least one solution is wrong, this tells us whether meson properties are useful tools to identify errors made in the determination of the correlation functions in the deep infrared.
 \item Finally, the hypothesis that scaling and decoupling are just gauge choices can be tested with such a setup: If they are, the meson properties may not depend on the choice, though this requires to take into account the gauge-dependence of the quark-gluon vertex, like in QED \cite{Curtis:1990zr,Kizilersu:2009kg}. However, this cannot verify the hypothesis; it can at most falsify it. If the hypothesis would be assumed to be correct, then the present study investigates how gauge-invariance is recovered in the Bethe-Salpeter equation, and a lower bound for the gauge-dependence of  the quark-gluon vertex is given.
\end{enumerate}
In any case, the main outcome of the present study will only be how sensitive some meson properties of some mesons are on the far infrared of the gauge-dependent correlation functions.

For the present purpose, all model parameters are fixed to pion properties; then the $\rho$-meson is studied.
Indeed, it turns out that the $\rho$ is essentially not affected by the choice of Yang-Mills solution, as long as the
quark-gluon vertex is adapted accordingly. This investigation complements similar ones, but with focus
on different observables \cite{Luecker:2009bs,Fischer:2009gk,Fischer:2009jm,Natale:2009uz,Aguilar:2004td}.

The article is organized as follows: The input scaling-type and decoupling-type solutions are detailed in section \ref{sym} and the
resulting quark propagator is studied in section \ref{squark}. The gauge-invariant results for the mesons are then
presented in section \ref{smeson}. A few concluding remarks can be found in section \ref{sconclusion}.

\section{Running gauge coupling}\label{sym}

As noted above, two sets of solutions for the Yang-Mills sector will be used in the present study, both of them
associated with the perturbative Landau gauge.

One is the so-called decoupling-type solution
\cite{Fischer:2008uz,Binosi:2009qm,Dudal:2009xh,Boucaud:2010gr,Cucchieri:2008fc,Bogolubsky:2009dc}. In this case,
the gluon propagator is infrared finite, thus exhibiting a screening mass, although it does not appear to have a pole mass
\cite{Fischer:2008uz,Cucchieri:2004mf,Bowman:2007du}. At the same time, the Faddeev-Popov ghost propagator is
that of an essentially massless particle. Correspondingly, also higher correlation functions are expected to
show a screening behavior although, except for some lattice studies in small volumes \cite{Cucchieri:2008qm}, this
has not been analyzed in detail.

The second solution is of the so-called scaling-type \cite{Fischer:2008uz,Zwanziger:2003cf,Lerche:2002ep}. It is
characterized by correlation functions showing an infrared power-law behavior \cite{Zwanziger:2003cf,Lerche:2002ep}.
All exponents are tied to one single base exponent, namely the one of the ghost propagator \cite{Alkofer:2004it,Fischer:2009tn}.
This propagator has to be more divergent than that of a massless particle. As a consequence, the gluon propagator has
to be either infrared finite, like in the decoupling solution, or infrared vanishing, depending on the value of this
critical exponent.

As already mentioned, is has been proposed that both solutions are just non-perturbative gauge choices
\cite{Fischer:2008uz,Maas:2008ri,Maas:2009se}, but so far it has not been possible to establish this beyond small lattice
volumes \cite{Maas:2009se}.

Irrespective of the solution, it is possible to define a running coupling by \cite{Fischer:2008uz,vonSmekal:1997is}
\be
\alpha(p^2)=\alpha(\mu^2)( G(p^2,\mu^2))^2 Z(p^2,\mu^2)\label{alpha}
\ee
\no in the so-called miniMOM scheme \cite{vonSmekal:2009ae}. Herein $G$ and $Z$ are the dimensionless ghost and gluon dressing
functions, respectively, which are obtained from the scalar part of the propagators by multiplication with $p^2$. For the
scaling-type solution this coupling is infrared finite,
while for the decoupling-type solution it is infrared vanishing like $p^2$. Note that this does not contradict the
conjecture that both could be just gauge choices, since a definition of a running coupling in terms of ghost fields is by construction not gauge-invariant \cite{vonSmekal:1997is}. However, if preferred, it is possible to translate this running coupling into the standard ${\overline\mathrm{MS}}$ scheme, at least perturbatively up to four loops \cite{vonSmekal:2009ae}. Nonetheless, for the present purpose, the form \pref{alpha} is much more convenient.

\begin{figure}
\includegraphics[width=\linewidth]{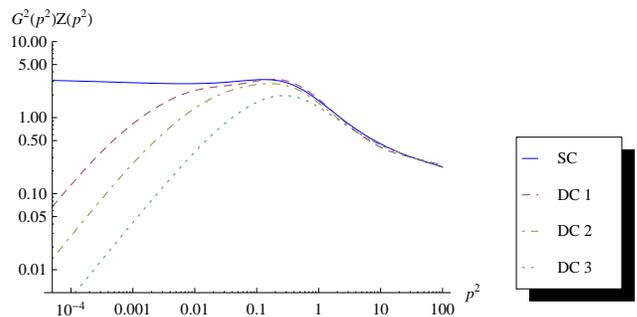}
\caption{\label{fig:alpha}The input running coupling \pref{alpha}. Here, and hereafter, SC denotes the scaling
solution and DC the decoupling solutions. For more details, see \cite{Fischer:2008uz}.}
\end{figure}

In the ladder truncation of the BSE coupled to the rainbow-truncated quark DSE, as we employ them
here, this running coupling is actually the only way the Yang-Mills sector enters, and the individual gluon and ghost
propagators are not needed. Here, the results for all solutions are taken from \cite{Fischer:2008uz}. Note that the
decoupling-type solution is possibly not unique \cite{Fischer:2008uz}, in agreement with small-volume lattice
simulations \cite{Maas:2009se}. For our purpose, we select three representatives, one of them (DC 3) being in rather good
quantitative agreement with lattice results \cite{Fischer:2008uz}, in order to study also the dependence
on different variants of decoupling-type behavior. We fit these numerical results with a simple ansatz, which faithfully
reproduces leading-order perturbation theory, along the lines of \cite{Fischer:2002hna}. The resulting input functions are
shown in figure \ref{fig:alpha}. Note that for simplicity we neglect the back-coupling of the quarks to the Yang-Mills
sector, which is only a rather weak effect \cite{Bowman:2007du}. It can be seen that the coupling quantitatively depends
on the choice of solution below a momentum scale of 1 GeV, and qualitatively below 100 MeV.

\section{Quarks}\label{squark}

In this context, a meson is treated as a quark-antiquark bound state via the homogeneous BSE \cite{LlewellynSmith:1969az}. In
such a setup, the dressed quark propagator $S(p)$ appears as an input in the BSE. For a quark with momentum $p$, the
propagator is given by
\be
S(p)=Z_q(p^2)\frac{-i\gamma_\mu p^\mu+M(p^2)}{p^2+M^2(p^2)}\nn,
\ee
with its two scalar dressing functions, the wave-function dressing $Z_q$ and the mass function $M$. To determine the
quark propagator, we employ the rainbow truncation of the quark DSE. The quark DSE or QCD gap equation
in its general form is given by
\begin{eqnarray}\nonumber
S(p)^{-1}  &=&  Z_2(i\gamma\cdot p + Z_m m_q)+  \Sigma(p)\,,\\\label{eq:gapequation}
\Sigma(p)&=& Z_1\, g^2\!\! \int^\Lambda_q\!\!\! D_{\mu\nu}(p-q)
\;\frac{\lambda^a}{2}\gamma_\mu \;S(q)\; \Gamma_\nu^a(p,q) \,,
\end{eqnarray}
where $\int^\Lambda_q$ denotes the integral $\int^\Lambda d^4q/(2\pi)^4$ regularized in a translationally invariant way with
the regularization scale $\Lambda$, $m_q$ is the current-quark mass at the renormalization scale, and the quark self energy $\Sigma$ involves
the renormalized dressed gluon propagator $D_{\mu\nu}$ and the renormalized dressed quark-gluon vertex
$\Gamma_\nu^a(p,q)$. Here, $a$ is a color index, $\lambda^a$ the Gell-Mann $SU(3)$ color matrices, $g$
the strong coupling constant, and the $Z_i$ are renormalization constants
(for more details and a full account of the renormalization procedure, see e.g.~\cite{Maris:1997tm}).

In rainbow truncation in Landau gauge the contributions from the dressed gluon propagator and the dressed
quark-gluon vertex are contracted to a single, effective object, for which we take the ansatz
\bea
\gamma_\mu \frac{\alpha(k^2)}{k^2}\left(1+4 \pi^2 k^2 \frac{D}{\omega^6} \exp^{-\frac{k^2}{\omega^2}}\right)
\left( \delta_{\mu\nu}-\frac{k_\mu k_\nu}{k^2} \right)\label{vertex}\\
=:\gamma_\mu \frac{\alpha(k^2)}{k^2} F(k^2,\omega,D)
\left( \delta_{\mu\nu}-\frac{k_\mu k_\nu}{k^2} \right).\label{eq:vertexdef}
\eea
The particular functional form of the effective interaction chosen here is inspired by various model ansaetze used in the
past which for the most part make use of the knowledge of the perturbative QCD running coupling as well as a simple
parametrizations of the low- and intermediate-momentum parts of the effective interaction
\cite{Jain:1991pk,Munczek:1991jb,Alkofer:2002bp,Maris:1997tm,Maris:1999nt}.
Also note that this has been done before in a similar way
in \cite{Fischer:2003rp}, albeit with a different ansatz for the vertex.

\begin{figure}
\includegraphics[width=\linewidth]{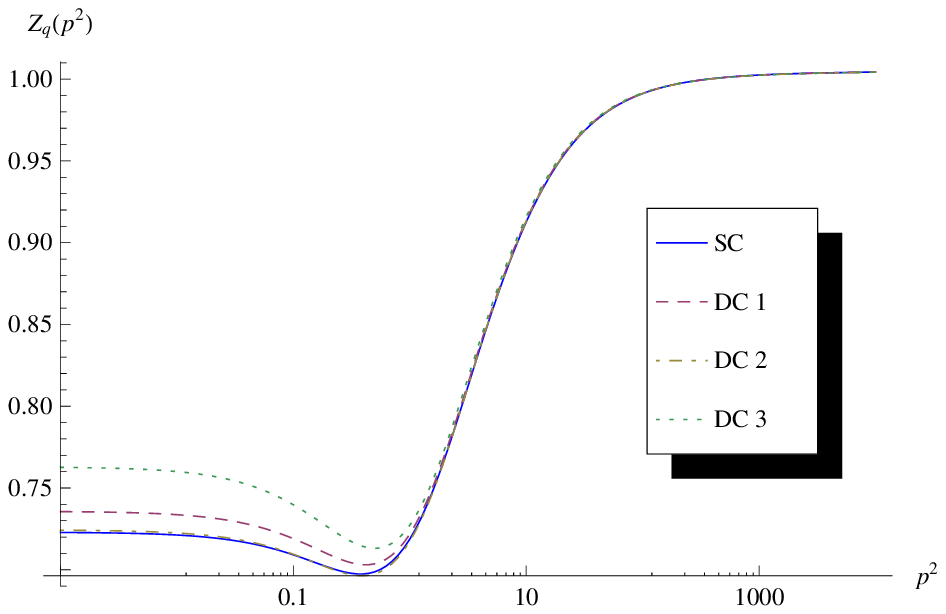}\\
\includegraphics[width=\linewidth]{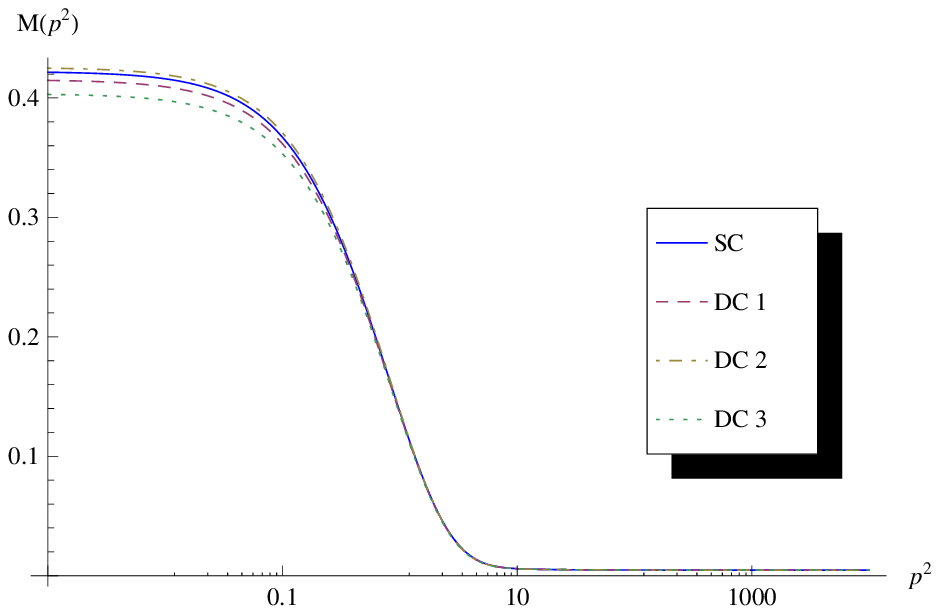}\\
\caption{\label{fig:quark}The quark wave-function dressing (top) and mass function (bottom) for the four choices of
parameters of the quark-gluon vertex fixed to pion properties, see table \ref{tab:fit}.}
\end{figure}

The parameters $D$ and $\omega$ are fixed by the pion properties as discussed below. At the same time, one has to fix
a value for the current quark mass which acts as a driving term for the non-perturbative $M$ in the quark DSE. Specified in
analogy to Ref.~\cite{Maris:1999nt} we have a value of $m_q=0.0047$ GeV at our renormalization point of $(19\mbox{ GeV})^2$ and
we use two degenerate flavors. The resulting quark dressing functions are shown in figure \ref{fig:quark}. They no
longer show any qualitative difference when comparing results corresponding to the scaling- and decoupling-type inputs; in fact,
only a small quantitative difference remains. The qualitative uniformity is a direct consequence of dynamical chiral symmetry
breaking, since it makes the quark effectively decouple in the infrared and thus blind to the qualitative differences.
The smallness of the quantitative differences among the solutions for $Z_q$ and $M$ turns out to stem from the requirement of
accurately reproducing the pion properties.

Note that some results are already available for the quark-gluon vertex and its further tensor structures as well
as its impact on bound-state calculations (see e.g. \cite{Alkofer:2008tt,Kizilersu:2006et,Matevosyan:2007cx,Williams:2009ce,Chang:2010xs}
and references therein). However, these are either specific to the scaling case \cite{Alkofer:2008tt}, from the lattice for
small volumes \cite{Kizilersu:2006et}, or from more involved calculations using functional methods
\cite{Matevosyan:2007cx,Williams:2009ce,Chang:2010xs} and cannot be used in the same way as the gluon input
for the comparison aimed at here.

\section{Mesons}\label{smeson}
In our study of meson properties, we employ the homogeneous quark-antiquark BSE, which in general form reads
\begin{eqnarray}\label{eq:generalbse}
\Gamma(k;P)&=&\int^\Lambda_q\!\!K(k;q;P)\;S(q_+) \Gamma(q;P) S(q_-)\,,
\end{eqnarray}
where $q_+$ and $q_-$ are the quark- and antiquark momenta, respectively. The Bethe-Salpeter amplitude $\Gamma(q;P)$ depends
on their total and relative momenta $P$ and $q$ with the relations $q_+=q+\eta P$ and $q_-=q-(1-\eta)P$.
The homogeneous BSE is only valid on-shell, i.e.~the total momentum of the bound state satisfies the (Euclidean) on-shell condition $P^2=-M^2$, where $M$ is the mass of the meson under consideration.
$\eta$ with $0\le \eta\le 1$ is the momentum partitioning parameter, which is arbitrary in the Lorentz-covariant context.
The 2-fermion irreducible quark-antiquark interaction kernel $K$ in general depends on the relative momenta as well as the total momentum
of the meson. In ladder truncation, the kernel does not depend on the total momentum, but only on the difference in relative
momentum (or in other words on the gluon momentum) $k-q$, and one obtains
\begin{eqnarray}\nonumber
\Gamma(k;P)&=&-\frac{4}{3}\int^\Lambda_q\!\frac{\alpha((k-q)^2)}{(k-q)^2}F((k-q)^2,\omega,D)\times \\\label{eq:bse}
&&\gamma_\mu^T \; S(q_+) \Gamma(q;P) S(q_-)\;\gamma_\nu  \,.
\end{eqnarray}
Note that the Dirac structure hidden in Eq.~(\ref{eq:generalbse}) has been made explicit through the Dirac $\gamma$ matrices
and the ordering of the terms under the integral has been chosen such that Dirac indices can be omitted. $\gamma^T$ denotes
the transversely projected $\gamma_\mu-\gamma_\tau (k_\tau-q_\tau)(k_\mu-q_\mu)/(k-q)^2$, and $\alpha$ and $F$ are the same
as in Eq.~(\ref{eq:vertexdef}).

As becomes clear from Eq.~(\ref{eq:generalbse}), the input quark propagator is needed for the arguments $q_\pm$, which are
complex vectors due to the on-shell condition. Therefore, an analytic continuation of the propagator is necessary. In this
context, a complication arises in the case of the scaling-type coupling. Since it does not go to zero as $p^2\rightarrow 0$,
it leads to a singularity in the vector part $\frac{1}{Z_q}$ of the inverse quark propagator \cite{Jarecke:2005ph,Eichmann:2009zx}.
Thus, in order to treat scaling- and decoupling-type inputs on the same footing, we employ the method detailed in
\cite{Krassnigg:2008gd} to obtain the solution of the gap equation in the complex plane. The method has two main features:
First, instead of the quark momentum as in Eq.~(\ref{eq:gapequation}), the gluon momentum is used as integration variable.
As a consequence, the coupling is only needed for real and positive arguments, whereas during the iterative solution
the quark propagator itself has to be evaluated for complex momenta, if the external momentum $p$ is complex. This, however,
leads to the second feature: By demanding that $\frac{1}{Z_q}$ and $\frac{M}{Z_q}$ are analytic in the domain of interest, the
gap equation can be iterated on the boundary of that region, and the values inside this boundary can be obtained in a numerically
	stable way from an appropriate implementation of the Cauchy formula \cite{Ioakimidis:1991io}. During the iteration, all singularities are
therefore eliminated and the result represents a solution of Eq.~(\ref{eq:gapequation}) in the complex plane which by construction
does not exhibit the kinematic singularity in the case of the scaling-type input, such that a consistent treatment of both types
of input is possible. Once the quark propagator is provided in such a way, the homogeneous BSE can be solved numerically
(for more details on the techniques used in this context see \cite{Blank:2010bp}).

Together with the rainbow truncation of the quark DSE, the ladder truncation of the BSE is used, since
this so-called rainbow-ladder truncation satisfies the axial-vector Ward-Takahashi identity in QCD, which encodes chiral symmetry
and related properties of the theory in the context of the DSEs \cite{Munczek:1994zz}. Satisfaction of this identity in
a model setup such as the one used
here guarantees the correct implementation of chiral symmetry and its dynamical breaking and produces, e.g., the well-known
result of a massless pion in the chiral limit. Via the identity one finds a generalized version of the Gell-Mann--Oakes--Renner
relation that is valid for every pseudoscalar meson, see \cite{Holl:2004fr} and references therein. With the pion
anchored in this way in such a calculation, its (and also kaon) properties have been subject to successful study for the past
years, exemplified by the excellent description of the pion's electromagnetic properties, e.g.~\cite{Maris:2000sk,Maris:2005tt}.
With similar success, vector meson properties have been investigated, e.g.~decay constants, hadronic decay widths, as well as
electromagnetic form factors and charge radii \cite{Maris:1999nt,Jarecke:2002xd,Bhagwat:2006pu}.

Thus, the $\pi$ and $\rho$ are ideal objects for the purpose of the present study. We present results for masses and decay
constants, where the latter are sensitive to the Bethe-Salpeter amplitude of the meson under consideration.
The numbers are compiled in two tables to highlight two different aspects of the investigation. First, we
keep the same vertex ansatz for all four different Yang-Mills input sets.
\begin{table}
\begin{center}
\begin{tabular}{l|c|c|c|c}
  &$m_\pi$& $f_\pi$&$m_\rho$&$f_\rho$\\
\hline
Scaling      & 0.1417&0.0932&0.7443&0.1492\\\hline
Decoupling 1 & 0.1420&0.0956&0.7641&0.1533\\
Decoupling 2 & 0.1401&0.0897&0.7155&0.1439\\
Decoupling 3 & 0.1389&0.0805&0.6321&0.1239\\
\hline
Experiment   & 0.1396&0.0922&0.7755&0.1527
\end{tabular}
\caption{\label{tab:unfit} Results for $\pi$ and $\rho$ mass and decay constant after adjusting the parameters
of the vertex ansatz \pref{vertex} such that the scaling solution best reproduces the pion properties. All quantities
are given in units of GeV; our results are accurate to less than half an MeV.}
\end{center}
\end{table}
We adjust the parameters of the vertex ansatz \pref{vertex} such that the scaling solution reproduces the pion mass
and decay constant adequately. The resulting values for $D$ and $\omega$ are 0.8 GeV${}^4$ and 0.8 GeV, respectively.
Then we compute the masses $m$ and decay constants $f$, $m_\pi$, $f_\pi$, $m_\rho$, and $f_\rho$ with the same parameters
also for the different decoupling-type cases; the corresponding results are listed in table \ref{tab:unfit} in units of GeV.
One can see that the more the input coupling differs from the scaling case, the larger the differences for $f_\pi$, $m_\rho$,
and $f_\rho$ become, while the general picture is not spoiled at a qualitative level. The small variation in $m_\pi$ is due
to the influence of the axial-vector Ward-Takahashi identity.
This increase in discrepancy when only changing the gluonic input has also been observed in the literature
\cite{Natale:2009uz,Aguilar:2004td}.

\begin{table*}
\begin{center}
\begin{tabular}{l|c|c|c|c|c|c}
  &$D$&$\omega$ &$m_\pi$& $f_\pi$&$m_\rho$&$f_\rho$\\
\hline
Scaling      &0.800&0.800& 0.1417&0.0932&0.7443&0.1492\\ \hline
Decoupling 1 &0.790&0.810& 0.1414&0.0932&0.7443&0.1496\\
Decoupling 2 &0.820&0.785& 0.1411&0.0936&0.7484&0.1502\\
Decoupling 3 &0.870&0.760& 0.1407&0.0935&0.7485&0.1514
\end{tabular}
\caption{\label{tab:fit}As table \ref{tab:unfit}, but $D$ and $\omega$ have now been adjusted for each input
separately to fit the pion's mass and decay constant. All quantities are given in units of GeV except for $D$
which has units of GeV${}^4$.}
\end{center}
\end{table*}

Next, we offer another possibility: irrespective of whether the different inputs correspond to different gauges
or are just different truncation assumptions, it is to be expected that the quark-gluon vertex should change
accordingly when the running coupling changes. To investigate this scenario, we have fitted the parameters $D$
and $\omega$ for each input individually to reproduce the pion parameters. The corresponding results are shown
in table \ref{tab:fit}. As one can clearly see, already a very moderate change in the parameters without
excessive finetuning yields agreement of $m_\pi$, $f_\pi$, $m_\rho$, and $f_\rho$ for all inputs on the percent level.
Thus, any consistent change of the quark-gluon vertex and the gluonic sector permits an equally good description of
$m_\pi$, $f_\pi$, $m_\rho$, and $f_\rho$. While not guaranteed, it is at least possible that a similar statement holds
for arbitrary other observables as well; some confirmation of our observation along these lines
has already been found in \cite{Luecker:2009bs,Fischer:2009jm}.

\begin{figure}
\includegraphics[width=\linewidth]{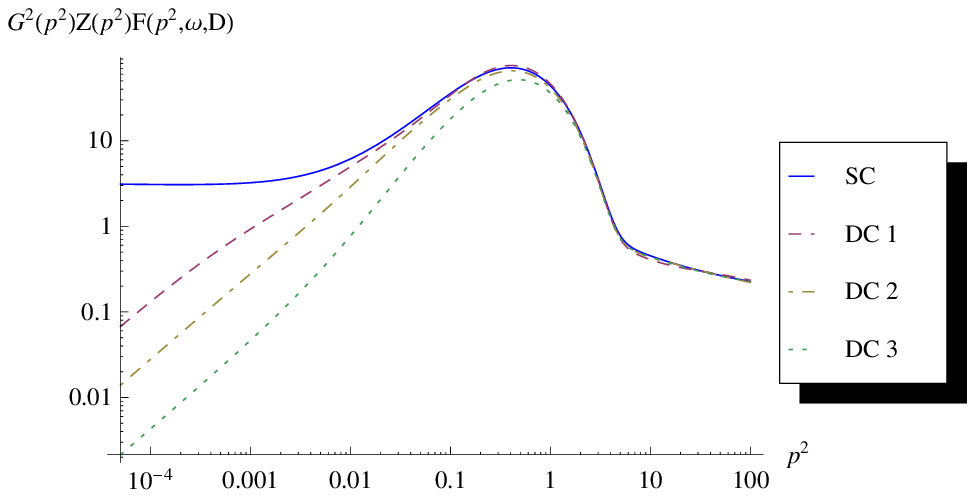}\\
\includegraphics[width=\linewidth]{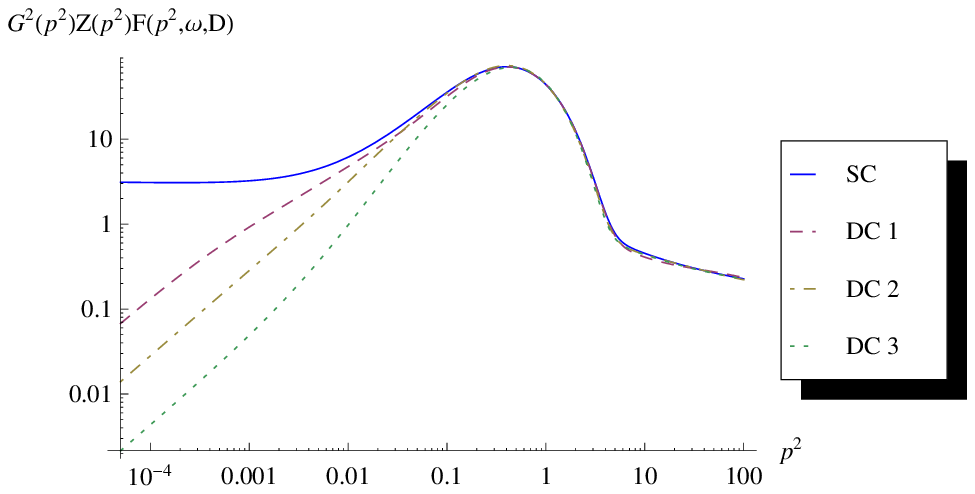}\\
\caption{\label{fig:ansatz}The momentum dependence of the ansatz \pref{vertex} for the parameter values $D$ and $\omega$ of table \ref{tab:unfit} (top) and for the optimized values of table \ref{tab:fit} (bottom).}
\end{figure}

It is interesting to check how the full ansatz \pref{vertex} looks for the optimized fits. More precisely, we
investigate the combined effect of the coupling and vertex dressing functions on different momentum scales.
This is shown in figure \ref{fig:ansatz}. It is nicely visible that the four cases for the ansatz start to
differ appreciably only below 1 GeV, but are qualitatively different in the far infrared. This observation supports
the argument that the $\rho$ has a gauge-invariant structure which exists at a level of 0.5 fm or
below \cite{Glozman:2009cp,Glozman:2010zn}, and is not too sensitive to larger scales.

\section{Conclusions}\label{sconclusion}

Summarizing, we have shown that in the coupled system of quark DSE and meson BSE in rainbow-ladder truncation,
changing the gluonic input in these equations can be compensated, at the level of the gauge-invariant hadronic properties,
by changing the corresponding quark-gluon vertex accordingly. This becomes apparent only at the level of the
gauge-invariant bound-state properties, in our case meson masses and decay constants, but not on the level of the
gauge-dependent quark propagator.

Returning to the three different questions posed in the introduction, this pattern in our results can be
interpreted as follows. For this two central observations are important. First, while the meson properties are qualitatively unchanged when changing some part of the input, there is some sensitivity to the far infrared left. Otherwise the results in table \ref{tab:unfit} would agree. Thus, the infrared is not irrelevant, and its behavior cannot be chosen arbitrarily. However, if all inputs are varied consistently, the meson properties are unchanged, see table \ref{tab:fit}, despite the fact that the infrared is still qualitatively different for the inputs, as is visible in figure \ref{fig:ansatz}. The reason for this insensitivity is found in the quark propagator, as discussed in section \ref{squark}: Chiral symmetry breaking screens the quark from the low-energy interactions, and thus makes it blind to the far infrared. The conclusions may thus be altered in the absence of chiral symmetry breaking, but then, the set of possible solutions in the gluon sector may differ as well. An indication of such a change is found in models with scalar instead of fermionic matter \cite{Maas:2010nc}.

This leads to the following conclusions:
\begin{enumerate}
\item[1.] From a technical point of view, it is not possible to deduce the infrared behavior of the input gauge-dependent correlation functions from the meson properties alone. Reproducing meson properties therefore not necessarily implies a faithful solution for the gauge-dependent correlation functions.
\item[2.] Though rather insensitive to the infrared, the $\rho$ meson is still influenced by the long-distance physics. Only a consistent long-distance behavior can reproduce the meson spectrum adequately.
\item[3.] If scaling and decoupling are indeed mere gauge choices, then, not surprisingly given the experience of QED \cite{Curtis:1990zr,Kizilersu:2009kg}, there is a gauge-dependence also in the matter sector. If this is taken into account, the results for mesons are gauge-invariant. However, in agreement with direct investigations \cite{Maas:2010nc}, the gauge-dependence in the matter sector is rather small.
\end{enumerate}
Thus in summary, the infrared region requires a consistent treatment when quantitative precise results are aimed at, but there is more than one infrared behavior leading to the same results. This is also an important result for the eventual aim of a self-consistent inclusion of the quark-gluon vertex in the future: Truncation artifacts in the far infrared are of smaller relevance than those at mid-momentum.

Regardless of this interpretation, we have reported results which can be taken as a guiding line for
how to proceed when aiming at getting gauge-invariant results from gauge-dependent correlation functions, as
supported also by other investigations \cite{Fischer:2009gk}. For this study, we have chosen a straight-forward, but
limited set of gauge-invariant observables, and we have made this clear together with the limits of the
interpretations provided, as appropriate. Naturally, further investigations along these or similar lines must
follow to shed more light on this important set of problems, which could be along the lines of corresponding QED investigations \cite{Curtis:1990zr,Kizilersu:2009kg}. A particular challenging and interesting topic would be Regge trajectories.\\

\no{\bf Acknowledgments}

We would like to thank R. Alkofer for helpful comments on the manuscript.
A.\ M.\ was supported by the Austrian Science Fund \emph{FWF} under grant M1099-N16.
M.~B.\ and A.~K.\ were supported by the \emph{FWF} under project no.\ P20496-N16.
The work of M.~B.\ has been performed in association with and supported in part by the
\emph{FWF} doctoral program no.\ W1203-N08.

\bibliographystyle{bibstyle}
\bibliography{had_nucl_graz}

\begin{thebibliography}{10}

\bibitem{Alkofer:2000wg}
R.~Alkofer and L.~von Smekal,
\newblock Phys. Rept. {\bf 353}, 281 (2001), hep-ph/0007355.

\bibitem{Fischer:2006ub}
C.~S. Fischer,
\newblock J. Phys. {\bf G32}, R253 (2006), hep-ph/0605173.

\bibitem{Roberts:2007jh}
C.~D. Roberts, M.~S. Bhagwat, A.~Holl, and S.~V. Wright,
\newblock Eur. Phys. J. ST {\bf 140}, 53 (2007), 0802.0217.

\bibitem{Krassnigg:2009zh}
A.~Krassnigg,
\newblock Phys. Rev. {\bf D80}, 114010 (2009), 0909.4016.

\bibitem{Eichmann:2009qa}
G.~Eichmann, R.~Alkofer, A.~Krassnigg, and D.~Nicmorus,
\newblock Phys. Rev. Lett. {\bf 104}, 201601 (2010), 0912.2246.

\bibitem{Eichmann:2007nn}
G.~Eichmann, A.~Krassnigg, M.~Schwinzerl, and R.~Alkofer,
\newblock Annals Phys. {\bf 323}, 2505 (2008), 0712.2666.

\bibitem{Nicmorus:2008vb}
D.~Nicmorus, G.~Eichmann, A.~Krassnigg, and R.~Alkofer,
\newblock Phys. Rev. {\bf D80}, 054028 (2009), 0812.1665.

\bibitem{Glozman:2009cp}
L.~Y. Glozman, C.~B. Lang, and M.~Limmer,
\newblock Few Body Syst. {\bf 47}, 91 (2010), 0909.2939.

\bibitem{Glozman:2010zn}
L.~Y. Glozman, C.~B. Lang, and M.~Limmer,
\newblock Phys. Rev. {\bf D82}, 097501 (2010), 1007.1346.

\bibitem{Bhagwat:2006pu}
M.~S. Bhagwat and P.~Maris,
\newblock Phys. Rev. {\bf C77}, 025203 (2008), nucl-th/0612069.

\bibitem{Fischer:2008uz}
C.~S. Fischer, A.~Maas, and J.~M. Pawlowski,
\newblock Annals Phys. {\bf 324}, 2408 (2009), 0810.1987.

\bibitem{Maas:2008ri}
A.~Maas,
\newblock Phys. Rev. {\bf D79}, 014505 (2009), 0808.3047.

\bibitem{Maas:2009se}
A.~Maas,
\newblock Phys. Lett. {\bf B689}, 107 (2010), 0907.5185.

\bibitem{Binosi:2009qm}
D.~Binosi and J.~Papavassiliou,
\newblock Phys. Rept. {\bf 479}, 1 (2009), 0909.2536.

\bibitem{Dudal:2009xh}
D.~Dudal, S.~P. Sorella, N.~Vandersickel, and H.~Verschelde,
\newblock Phys. Rev. {\bf D79}, 121701 (2009), 0904.0641.

\bibitem{Boucaud:2010gr}
P.~Boucaud {\em et~al.},
\newblock (2010), 1004.4135.

\bibitem{Zwanziger:2003cf}
D.~Zwanziger,
\newblock Phys. Rev. {\bf D69}, 016002 (2004), hep-ph/0303028.

\bibitem{Lerche:2002ep}
C.~Lerche and L.~von Smekal,
\newblock Phys. Rev. {\bf D65}, 125006 (2002), hep-ph/0202194.

\bibitem{Alkofer:2004it}
R.~Alkofer, C.~S. Fischer, and F.~J. Llanes-Estrada,
\newblock Phys. Lett. {\bf B611}, 279 (2005), hep-th/0412330.

\bibitem{Fischer:2009tn}
C.~S. Fischer and J.~M. Pawlowski,
\newblock Phys. Rev. {\bf D80}, 025023 (2009), 0903.2193.

\bibitem{Fischer:2007pf}
C.~S. Fischer, A.~Maas, J.~M. Pawlowski, and L.~von Smekal,
\newblock Annals Phys. {\bf 322}, 2916 (2007), hep-ph/0701050.

\bibitem{Fischer:2005ui}
C.~S. Fischer, B.~Gr{\"u}ter, and R.~Alkofer,
\newblock Ann. Phys. {\bf 321}, 1918 (2006), hep-ph/0506053.

\bibitem{Cucchieri:2008fc}
A.~Cucchieri and T.~Mendes,
\newblock Phys. Rev. {\bf D78}, 094503 (2008), 0804.2371.

\bibitem{Bogolubsky:2009dc}
I.~L. Bogolubsky, E.~M. Ilgenfritz, M.~M{\"u}ller-Preussker, and A.~Sternbeck,
\newblock Phys. Lett. {\bf B676}, 69 (2009), 0901.0736.

\bibitem{Kizilersu:2009kg}
A.~Kizilersu and M.~R. Pennington,
\newblock Phys. Rev. {\bf D79}, 125020 (2009), 0904.3483.

\bibitem{Curtis:1990zr}
D.~C. Curtis, M.~R. Pennington, and D.~A. Walsh,
\newblock Phys. Lett. {\bf B249}, 528 (1990).

\bibitem{Luecker:2009bs}
J.~Luecker, C.~S. Fischer, and R.~Williams,
\newblock Phys. Rev. {\bf D81}, 094005 (2010), 0912.3686.

\bibitem{Fischer:2009gk}
C.~S. Fischer and J.~A. Mueller,
\newblock Phys. Rev. {\bf D80}, 074029 (2009), 0908.0007.

\bibitem{Fischer:2009jm}
C.~S. Fischer and R.~Williams,
\newblock Phys. Rev. Lett. {\bf 103}, 122001 (2009), 0905.2291.

\bibitem{Natale:2009uz}
A.~A. Natale,
\newblock PoS {\bf QCD-TNT09}, 031 (2009), 0910.5689.

\bibitem{Aguilar:2004td}
A.~C. Aguilar, A.~Mihara, and A.~A. Natale,
\newblock Int. J. Mod. Phys. {\bf A19}, 249 (2004).

\bibitem{Cucchieri:2004mf}
A.~Cucchieri, T.~Mendes, and A.~R. Taurines,
\newblock Phys. Rev. {\bf D71}, 051902 (2005), hep-lat/0406020.

\bibitem{Bowman:2007du}
P.~O. Bowman {\em et~al.},
\newblock Phys. Rev. {\bf D76}, 094505 (2007), hep-lat/0703022.

\bibitem{Cucchieri:2008qm}
A.~Cucchieri, A.~Maas, and T.~Mendes,
\newblock Phys. Rev. {\bf D77}, 094510 (2008), 0803.1798.

\bibitem{vonSmekal:1997is}
L.~von Smekal, R.~Alkofer, and A.~Hauck,
\newblock Phys. Rev. Lett. {\bf 79}, 3591 (1997), hep-ph/9705242.

\bibitem{vonSmekal:2009ae}
L.~von Smekal, K.~Maltman, and A.~Sternbeck,
\newblock Phys. Lett. {\bf B681}, 336 (2009), 0903.1696.

\bibitem{Fischer:2002hna}
C.~S. Fischer and R.~Alkofer,
\newblock Phys. Lett. {\bf B536}, 177 (2002), hep-ph/0202202.

\bibitem{LlewellynSmith:1969az}
C.~H. Llewellyn-Smith,
\newblock Ann. Phys. {\bf 53}, 521 (1969).

\bibitem{Maris:1997tm}
P.~Maris and C.~D. Roberts,
\newblock Phys. Rev. {\bf C56}, 3369 (1997), nucl-th/9708029.

\bibitem{Jain:1991pk}
P.~Jain and H.~J. Munczek,
\newblock Phys. Rev. {\bf D44}, 1873 (1991).

\bibitem{Munczek:1991jb}
H.~J. Munczek and P.~Jain,
\newblock Phys. Rev. {\bf D46}, 438 (1992).

\bibitem{Alkofer:2002bp}
R.~Alkofer, P.~Watson, and H.~Weigel,
\newblock Phys. Rev. {\bf D65}, 094026 (2002), hep-ph/0202053.

\bibitem{Maris:1999nt}
P.~Maris and P.~C. Tandy,
\newblock Phys. Rev. {\bf C60}, 055214 (1999), nucl-th/9905056.

\bibitem{Fischer:2003rp}
C.~S. Fischer and R.~Alkofer,
\newblock Phys. Rev. {\bf D67}, 094020 (2003), hep-ph/0301094.

\bibitem{Alkofer:2008tt}
R.~Alkofer, C.~S. Fischer, F.~J. Llanes-Estrada, and K.~Schwenzer,
\newblock Annals Phys. {\bf 324}, 106 (2009), 0804.3042.

\bibitem{Kizilersu:2006et}
A.~Kizilersu, D.~B. Leinweber, J.-I. Skullerud, and A.~G. Williams,
\newblock Eur. Phys. J. {\bf C50}, 871 (2007), hep-lat/0610078.

\bibitem{Matevosyan:2007cx}
H.~H. Matevosyan, A.~W. Thomas, and P.~C. Tandy,
\newblock J. Phys. {\bf G34}, 2153 (2007), 0706.2393.

\bibitem{Williams:2009ce}
R.~Williams and C.~S. Fischer,
\newblock (2009), 0912.3711.

\bibitem{Chang:2010xs}
L.~Chang and C.~D. Roberts,
\newblock AIP Conf. Proc. {\bf 1261}, 25 (2010), 1004.1848.

\bibitem{Eichmann:2009zx}
G.~Eichmann,
\newblock {\em {Hadron properties from QCD bound-state equations}},
\newblock PhD thesis, Karl-Franzens-University Graz, 2009, 0909.0703.

\bibitem{Jarecke:2005ph}
D.~W. Jarecke,
\newblock {\em Properties of mesons from Bethe-Salpeter amplitudes},
\newblock PhD thesis, Kent State University, 2005,
\newblock Bethe-Salpeter; pion; rho.

\bibitem{Krassnigg:2008gd}
A.~Krassnigg,
\newblock PoS {\bf CONFINEMENT8}, 075 (2008), 0812.3073.

\bibitem{Ioakimidis:1991io}
N.~I. Ioakimidis, K.~E. Papadakis, and E.~A. Perdios,
\newblock BIT Numerical Mathematics {\bf 31}, 276 (1991).

\bibitem{Blank:2010bp}
M.~Blank and A.~Krassnigg,
\newblock (2010), 1009.1535.

\bibitem{Munczek:1994zz}
H.~J. Munczek,
\newblock Phys. Rev. {\bf D52}, 4736 (1995), hep-th/9411239.

\bibitem{Holl:2004fr}
A.~Holl, A.~Krassnigg, and C.~D. Roberts,
\newblock Phys. Rev. {\bf C70}, 042203 (2004), nucl-th/0406030.

\bibitem{Maris:2000sk}
P.~Maris and P.~C. Tandy,
\newblock Phys. Rev. {\bf C62}, 055204 (2000), nucl-th/0005015.

\bibitem{Maris:2005tt}
P.~Maris and P.~C. Tandy,
\newblock Nucl. Phys. Proc. Suppl. {\bf 161}, 136 (2006), nucl-th/0511017.

\bibitem{Jarecke:2002xd}
D.~Jarecke, P.~Maris, and P.~C. Tandy,
\newblock Phys. Rev. {\bf C67}, 035202 (2003), nucl-th/0208019.

\bibitem{Maas:2010nc}
A.~Maas,
\newblock Eur. Phys. J. C in print  (2010), 1007.0729.

\end{thebibliography}


\end{document}